# Compact 3 DOF Driving Simulator using Immersive Virtual Reality

Jungsu Pak[2,4], Uri Maoz[1,2,3,4,5,6]

1) Crean College of Health and Behavioral Sciences at Chapman University
2) Schmid College of Science and Technology at Chapman University
3) Fowler School of Engineering at Chapman University
4) Institute for Interdisciplinary Brain and Behavioral Sciences at Chapman University
5) Anderson School of Management University of California Los Angeles
6) Biology and Bioengineering at California Institute of Technology

**Abstract**

A driving simulator was created using commercially available 3 degree of freedom motion platform (DOFreality H3) and a virtual reality head-mounted display (Oculus CV1). Using virtual reality headset as the visual simulation system with low-cost moving base platform allowed us to create a high-fidelity driving simulator with minimal cost and space. A custom motion cueing algorithm was used to minimize visuo-vestibular conflict, and simulator sickness questionnaire (SSQ) was used to measure progression of simulator sickness over time while driving on a highway environment.

**Introduction**

From the perspective of cognitive psychology, driving requires a host of cognitive and decision-making skills in a wide range of scenarios [1,2,3]. Driving simulators (DSs) allow researchers to create such driving scenarios in a safe and well controlled environment. Such scenarios might range from simply following another car to a morally difficult dilemma where a fatal accident is inevitable and choices must be made about mitigating harm [4,5]. State-of-art driving simulators often consist of three components [6,7]

1. Dome/cylindrical projection as the visual system
2. High degree-of-freedom (DOF) motion platform, and
3. Physical model of a vehicle or cockpit.

For those components to work well together, the vehicle model must be positioned at a distance from a wide projection screen to provide a proper visual field, while being on top of a bulky moving base for the rotational DOF through simultaneous rotation of all three components. For simulators with higher DOF, the moving base may sit on a rail system, where the translational DOF is simulated by sliding the entire setup along the rail. While such combination of expensive and bulky equipment may improve the performance of the driving simulator, the cost and space requirements for such a setup considerable.

Several teams have tried using virtual reality (VR) head mounted displays (HMDs) as the visual system of a driving simulators to minimize its size, space, and cost [8,9,10,11]. An HMD utilizes its binocular display and head tracking to fully immerse a participant into a virtual environment, where the vehicle model and driving environment are simulated. This allows significant reduction in the space requirements and thus in cost. Few teams tried to create a driving simulator with an HMD and a moving base [8]; the majority of implementation did not include such a base [9,10,11]. The latter teams have reported varying—though often considerable—degrees of simulator sickness (SS), characterized by nausea, headache, and fatigue, after participants' exposure to the virtual driving environment. This has made SS a prominent obstacle for using HMDs in driving simulation.

Sensory-conflict theory is a predominant explanation for SS [12, 13], having been reviewed and expanded for decades. The theory argues that incongruent stimuli between the visual and vestibular systems causes SS. When using only an HMD in a DS, visuo-vestibular conflict occurs when visual stimuli suggest that a force is acting on the subject (e.g., when turning and accelerating/decelerating) while the vestibular system does not register that force. The true potential of an HMD for DSs can therefore be realized only when it is paired with a moving base, where the HMD and the moving base provide synchronized, congruent stimuli and minimize the SS.

Here we describe a compact driving simulator that we created by integrating an HMD (Oculus Rift CV1) with a 3 DOF moving base (DOFreality H3), and a simulation environment programmed in Unity 3D. A custom motion cueing algorithm was used to maintain the synchrony of visual stimuli and vestibular stimuli to minimize SS as sensory conflict theory suggests. We describe the system below as well as a pilot experiment that we conducted to test how well our HMD and moving platform helped alleviate SS.

**Hardware**

Our driving simulator (Fig. 1) consists of three parts:

1. The visual system—Oculus Rift CV1.
2. Steering wheel, pedals, and shift gear—Logitech G920, and
3. Motion platform with 3 DOFs—DOFreality H3

DOFreality H3 is a commercial moving base platform that utilizes three motors that rotate up to 20° at the speed of 80 degrees per second to provide pitch, roll, and yaw. Two frontal motors are connected to a frontal axle and rotate in same direction to create pitch range of -4.4 to 6.6 degrees and rotate in opposite directions to create roll range of -9.0 to 9.0 degrees. The third motor, located at the rear side of the platform, rotates the seat and the two frontal motors to create yaw range of -10.0 to 10.0 degrees. The Logitech G920 is the input device of the DS and consists of a motorized steering wheel, pedals, and a shift. The wheel rotation range is -450 to 450 degrees. It is also capable of force feedback, such as vibrations and rotational forces. The Oculus Rift CV1 is a VR HMD used as the visual system, providing rotational and positional tracking of the head of the participant. The sensor for Oculus Rift, the Logitech G920 controls, and the seat and slider are mounted on solid metal frames provided by DOFreality H3. The total dimensions of the driving simulator, excluding the power unit for the moving base and the computing unit, is 91x152x61 cm (3x5x2 ft).

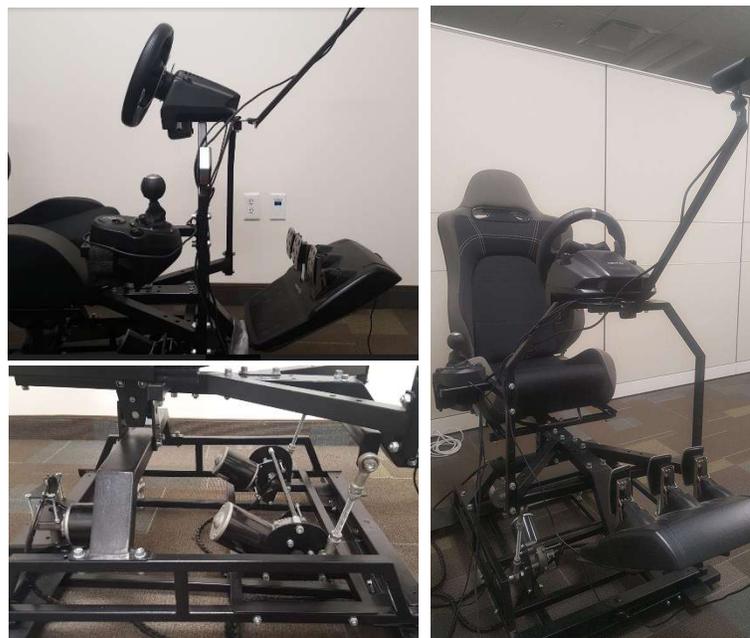

*Figure 1. Images of driving simulator without power unit and the computing unit. Left top shows Logitech G920; Left bottom shows DOFreality H3; Right shows the entire setup.*

**Software**

The simulation environment was created in Unity3D (Unity Technologies) using VR assets acquired from the Unity asset store (3D models, materials, and scripts). Unity3D was chosen for its compatibility with the Oculus Rift and with the physics simulation of the Nvidia PhysX 3.3. Inputs from the Logitech G920 components was used as input to the Nvidia PhysX 3.3, which is the default physics simulation for Unity3D. The result of the physics simulation was converted to inputs for DOFreality H3 using custom C# codes. We used Simtools as a middleware that connected C# codes to the moving base with professional license obtained from the purchase of DOFreality H3. The C# script from the Unity Standard Asset package vehicle was modified to allow participants to control the virtual vehicle using the Logitech G920, which was farther programmed to vibrate. Vibration frequency and magnitude were programmed to fluctuate with the velocity of the vehicle. The shift of Logitech G920 resembles that for manual transmission, but was used to shift between drive, neutral, and reverse only, as the virtual vehicle was programmed to operate as with an automatic transmission. For the same reason, the leftmost pedal of the three pedals—typically serving as the clutch—was not used for our automatic-transmission vehicle. SMC3Utils.exe was a software obtained from DOFreality to farther optimize the parameters of servos and was used to change the parameters of each servos such as power usage, dead zone, clipping angles, smoothing, and extras.

**Motion-Cueing Algorithm**

Conventional motion-cueing algorithms for 3 DOF simulators typically utilize low-pass and high-pass filters on the acceleration to smooth out jerky movements and break those down to simulate translation and rotation, where translation is expressed through rotating at a subthreshold rate to a position where gravity is used to simulate the force [14,15,16]. However, such smoothing algorithms requires sampling from multiple instances of the acceleration and introduce incongruency between visual stimuli and vestibular stimuli and hence exacerbate SS. In order to minimize the incongruency, the low-pass filter was applied to the throttle and brake pedals' input before they were fed into physics simulation as shown in Fig 2. This worked as a driving-assistance mechanism, where jerky movements from abrupt change in the input were dampened. The simulated values from the physics simulation—such as Y angular velocity and XZ plane rotation—could be directly accessed while each acceleration in the X or Z directions was calculated through the change of velocity after each frame. The centrifugal force was calculated using acceleration, velocity, and angular velocity of the simulated vehicle; and was expressed as roll. The simulated values for X and Z linear accelerations were scaled and applied to pitch and roll without filtering, respectively. A typical washout filter was applied for yaw, where high pass filter was applied to Y angular velocity and subthreshold movement towards the origin calculated yaw.

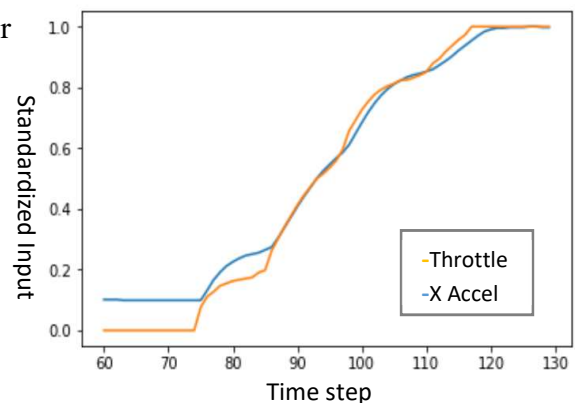

*Figure 2. First order low pass filter was applied to throttle. The filtered throttle was added to a constant idle acceleration based on while shift was in "Drive".*

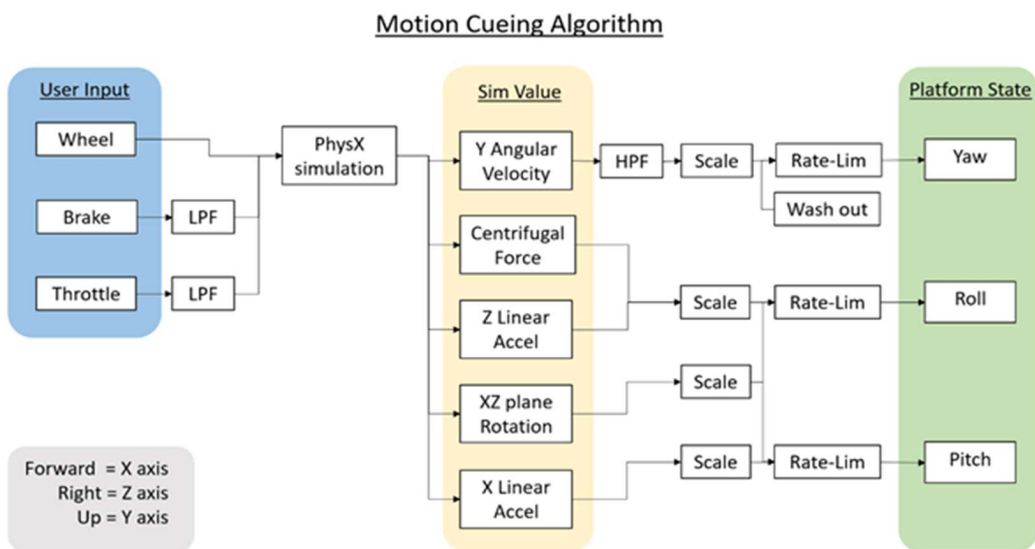

Figure 3. Diagram of the custom motion cueing algorithm. The simulated inputs were not filtered as rendered images were based on the exact value of the simulation. Instead, first order low pass filter was applied to throttle and brake inputs before feed

**Synchrony**

In order to synchronize the execution of visual and vestibular stimuli, Unity3D's default computing pipeline was tweaked because the default configuration of execution introduces delay between vestibular and visual stimuli. The default main steps in the order of execution of the Unity-C# loop are the following: *FixedUpdate*, *Update*, and *Rendering* [17]. For ease of explanation, we go through the timeline in reverse. *Rendering* is the final phase where shapes, highlights, and shadows are drawn using DirectX to create a frame of the visual display after the *Update*. The *Update* is the main function that runs once per frame, after *FixedUpdate* and before the *Rendering*, and is where codes for input processing and logical operations occur before they are drawn in *Rendering*. *FixedUpdate* is used to simulate running a body of code every fixed amount of time. It simulates by getting the time it took for the previous frame to execute and running the function according to the amount of times it would have ran if it were running in parallel. If the time it spent creating the previous frame was shorter than the predetermined fixed amount of time, *FixedUpdate* would skip that frame. By default, the physics simulation is executed at the end of *FixedUpdate*, and most of codes for the physics should go into *FixedUpdate*. However, this leads to inconsistent physical simulation, which leads to visuo-vestibular conflict and enhances SS.

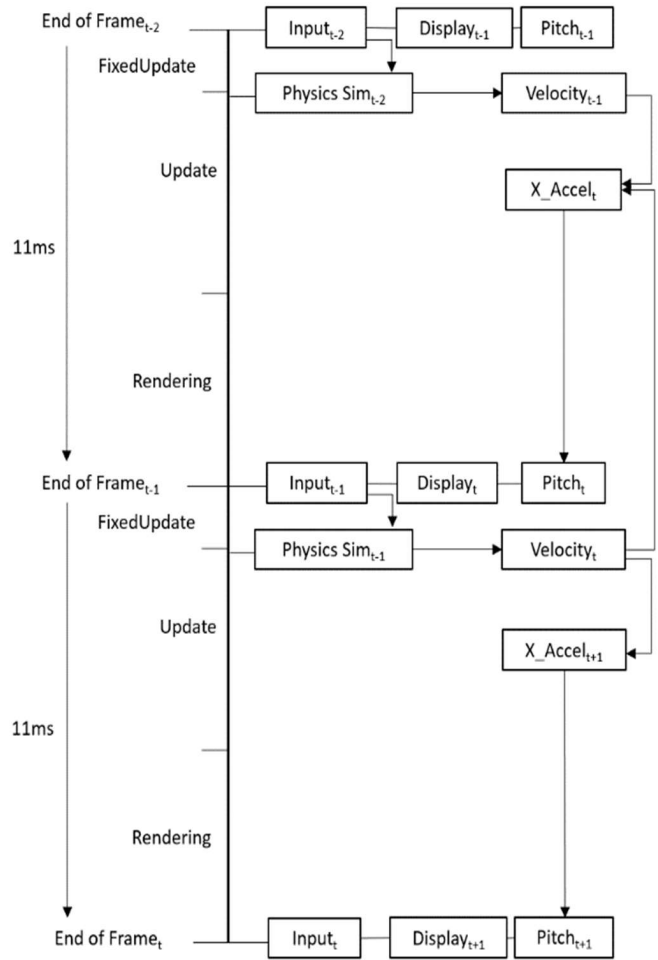

The automatic physical simulation done by Unity in the *FixedUpdate* was therefore disabled and was manually done at the beginning of the *Update* step to ensure that each frame will be accompanied by the most recent result of the physics simulation. This method was also used together with another technique, called Vertical Synchronization (VSync), which keeps the frame rate synchronized with the display frequency to reduce computing unnecessary frames that will not be rendered due to the limitation of the display frequency of the monitor. As the display frequency of the Oculus Rift is 90 Hz, a single frame buffer results in approximately 11 ms delay between frames. A single frame buffer was introduced through the Nvidia graphics driver setting to match the displayed frame with the proper pitch and roll, calculated by differencing velocity. This buffer introduced an input delay of 11 ms. But it also enabled computing the next frame while the current frame was being displayed, resulting in a smoother display.

*Figure 4. Simplified timeline of Unity3D showing the computation of pitch. Display and pitch are matched in time step. Input was always one time step behind, introducing a input latency of 11ms.*

**Simulator Sickness**

As discussed above, SS is a major impediment to using HMDs in DSs and enjoying the cost, space, and emersion benefits that such systems offer. A key concern for us when building our DS was therefore to minimize SS. And here we report the results of a pilot experiment that we carried out to test SS with our system.

Six participants (female = 5, male =1, mean age = 24.5, std =2.14) were recruited to validate the usability of our DS in terms of SS. Participants drove on a 4-lane highway that consists of 9 turns: {30, 60, 90} degrees x slope of {incline, decline, plateau} at 100 meters radius as shown in Figure 5. Participants were given 2 minutes to get accustomed to the virtual reality vehicle and the simulator. During this time, the vehicle was idle, and the instructions regarding driving were given by the experimenter. Participants were instructed to carefully drive on the left lane of the road and avoid crashes. After each lap (mean = 3.44 minutes, std = 0.94 minutes), the Simulator Sickness Questionnaire (SSQ) [18] was administered verbally to participants for a total of 6 laps (mean = 27.04 minutes, std = 2.70). The trials terminated when participants crashed, and SSQ was given immediately. After the questionnaire of the last lap, participants were instructed to take off the headset, step outside of DS, stand on two feet facing the experimenter, and answer the last SSQ. The SSQ scores of all participants are illustrated in Fig. 6. As is apparent, there is an upward trend on SS scores for some participants while flat line for others. On Average, there was a trend of 1.44 total score increase per trial.

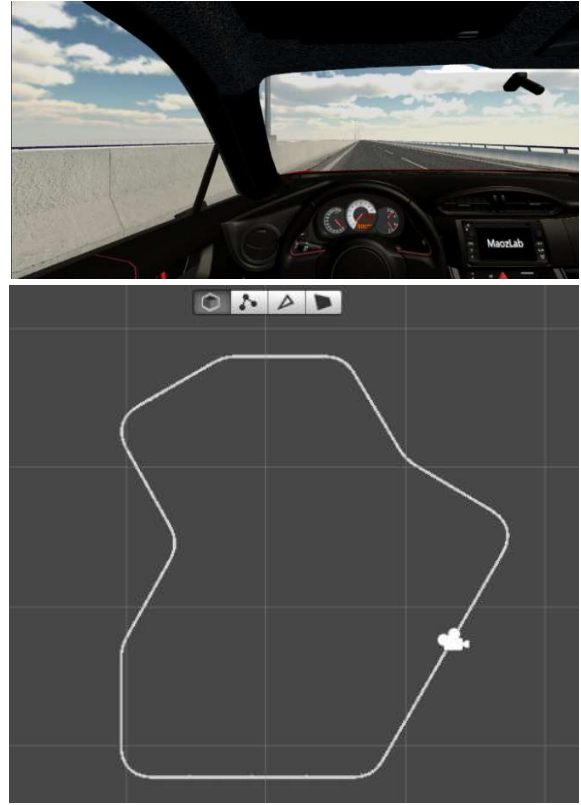

Figure 5. Top shows the Left Eye view of participant. The mirrors were disabled for consistency of rendering frame rate. The bottom shows the track that participants drove. The camera icon indicates the start and finish line.

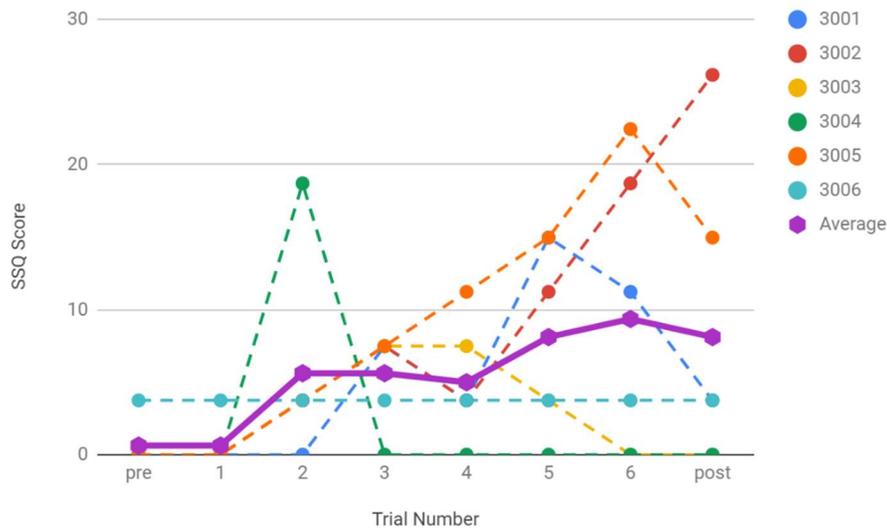

Figure 6. Total score of SSQ for each participant and on average. There seems to be clear trend up, but the overall severity of SS was minimal.

## Conclusion

The quality of DSs ranges from fixed-base simulators with a single monitor to a whole warehouse rigged with systems of rails and dome projections. The driving simulator we describe here is unique in three aspects: cost, space, and flexibility—i.e., it is relatively inexpensive, takes up little space, and is highly flexible and immersive. We took a great effort in providing synchronized visual and vestibular stimuli to the participants and measured their SSQ over time. As expected, the overall trend of SSQ increased over time, but the score after approximately 30 minutes of exposure was still found to be minimal. This suggests either that even the very minute visuo-vestibular conflict can increase SSQ over time, or that there is another cause of SSQ outside of sensory conflict model as other theories of SS in literature suggests. What is more, for 3 of the 6 participants, the SSQ score either remained flat or rose and then decreased back to 0 with time, suggesting that the cause of SSQ may be a binary trait. Needless to say, the number of participants collected here is much too small to generalize from our results. And this study should be viewed as preliminary. But it suggests that it is possible to create an HMD-based DS where the effects of SS are mitigated enough to allow participants to drive for about 30 minutes without suffering from serious SS. That is enough time to run various decision-making, perceptual, and other experiments on the DS.

There is much room for improvement in our DS. On the hardware side, the servos moving the base could be upgraded to make it more responsive and with a wider range of movement and thus with higher force ratio. The center of rotation can also be improved by moving it above the user, which would allow the expression of rotation as well as translation through rotation to be in the same direction. Having an HMD as visual system is important for this, as it facilitates making the DS compact and lightweight enough to be ceiling mounted. With these changes, better motion-cueing algorithms can be used. The motion cueing algorithm used here was rudimentary.

Another direction for the improvement would be the HMD itself. Current refresh rates of the monitor of the Oculus CV1 is 90Hz, which enforces an 11 ms delay between frames. The synchronization method we used for simulation to match the visual and vestibular stimuli mandated a one-frame delay. A higher refresh-rate monitor—e.g. 140Hz—would reduce that delay—to 7 ms, respectively—while maintaining the synchrony of stimuli. Low overall latency is especially important in high-speed driving where the vehicle's position changes rapidly. Increasing the field of view is also a desired feature. Peripheral vision is an important component in human perception of self-motion [19,20].Some early participants (not included here) commented that they did not experience going as fast as the speedometer noted, before the tactile stimulus, in the form of steering-wheel vibrations, was introduced to indicate the speed.

The higher refresh rate and the field of view naturally would require a stronger computing power. Unity3D is a commercial game-engine where high-quality visual effects and environment can be created. The higher computing power would not only make the frame rate steady, but also allow more enriched environment with cutting edge visual effects and realisms.

The above clearly shows how much VR-dependent research is dependent on technological advances and especially stronger computational power. However, as we also show above, good DS results can be achieved also for relatively inexpensive systems.